\documentclass[a4paper]{jpconf}
\usepackage{graphicx}

\usepackage{color}
\usepackage{graphicx}

\usepackage{amsfonts}
\def\be{\nopagebreak[3]\begin{equation}}
\def\ee{\end{equation}}
\def\ba{\nopagebreak[3]\begin{eqnarray}}
\def\ea{\end{eqnarray}}
\def\d{{\rm d}}
\def\b{{\beta}}

\def\d{{\rm d}}

\def\co{\sqrt{12 \pi G}}

\def\Tr{{\rm Tr\,}}

\def\g{\gamma}
\def\lp{{\ell}_{\rm Pl}}

\def\q{\mathring{q}}
\def\e{\mathring{e}}
\def\ow{\mathring{\omega}}
\def\ov{V_o}

\newcommand{\rcr}{\rho_{\mathrm{crit}}}

\newcommand{\p}{\partial}

\newcommand{\f}{\frac}

\def\f{\frac}

\def\ul{\underline}

\def\t{\tilde}

\begin{document}
\title{Transcending Big Bang in Loop Quantum Cosmology: Recent Advances\footnote{Based on the plenary talk in the Sixth International Conference on Gravitation \& Cosmology, December 17-21, 2007 at the Inter-University Centre for Astronomy and Astrophysics, Pune.}}
\author{Parampreet Singh}
\address{Perimeter Institute for Theoretical Physics, 31 Caroline Street North,\\ Waterloo, Ontario N2L 2Y5, Canada.}
\ead{psingh@perimeterinstitute.ca}

\begin{abstract}

We discuss the way 
non-perturbative quantization of cosmological spacetimes in loop quantum cosmology provides 
insights on the physics of Planck scale and the resolution of big bang singularity.
In recent years, rigorous examination of mathematical and physical aspects 
of the quantum theory has led to a consistent quantization which is consistent and physically viable and some early  ideas have been ruled out. The latter include so called `physical effects' originating from modifications to inverse scale factors in the flat models. The singularity resolution is understood to originate from the non-local nature of curvature in the quantum theory and the underlying polymer representation.  Using an exactly solvable model various insights have been gained. 
The model predicts a generic occurrence of bounce for states in the physical Hilbert space and a supremum for the spectrum of the energy density operator. It also 
provides answers to the growth of fluctuations, showing that  semi-classicality is preserved to an amazing degree across the bounce.
\end{abstract}

\section{Introduction}

Big bang is conventionally associated in cosmology as the beginning of space and time. However, it is an event beyond the realm of general relativity (GR). As the scale factor approaches zero, the energy density and spacetime curvature diverge and the evolution breaks down. The occurrence of singularity signals that a more fundamental theory should provide a description at these scales. Loop quantum gravity (LQG) is a non-perturbative and background independent quantization of GR \cite{lqg} whose predictions include a discrete quantum geometry underlying the classical continuum spacetime. It has been successfully applied to understand  aspects of 
black hole thermodynamics \cite{lqg_bh} and in recent years considerable progress has been made to compare non-perturbative loop quantization with conventional perturbative schemes and insights have been obtained to derive the graviton propagator \cite{sf}. A powerful result
originating from the background independence is the uniqueness of kinematical representation of the quantum theory which forms the basis of the various novel predictions.

Lessons from loop quantum gravity have been applied in simple models resulting 
from symmetry reduction. In cosmological spacetimes, given the underlying symmetries the quantization program can be completed and physical predictions can be extracted. In this approach, known as loop quantum cosmology (LQC), one uses 
the methods and techniques developed in LQG \cite{aarev,abl,mb1}. The strategy is to cast the 
classical phase space in Ashtekar variables and use holonomies of connection and fluxes of the triad as the elementary variables for quantization. The resulting quantum theory turns out to be in-equivalent to the Wheeler-DeWitt quantization.
The discreteness of underlying quantum geometry plays a fundamental role to provide novel physics at Planck scale resulting in resolution of 
big bang singularity and the occurrence of a quantum bounce at the Planck scale when energy density reaches a critical value \cite{apslett,aps2}. These results have been established for massless scalar field with/without a cosmological constant in a flat, closed and open topologies \cite{aps1,warsaw,apsv,open}. Investigations of models with massive scalar field reveal similar features \cite{aps3}.
Through a recently developed exactly solvable model \cite{acs}, solvable LQC (sLQC), 
a greater understanding has been obtained on the physical predictions.

We will focus on the flat isotropic model and start with a discussion of the classical phase space in the Ashtekar variables and the way it is related to usual spacetime description.
We will then move to the kinematical aspects of quantization and discuss the way different terms in the classical constraint are quantized. This will be followed by the resulting dynamics from LQC. Though a major part of the discussion will be on new improved dynamics of LQC 
which is singled out to be physically viable from a large class of in-equivalent quantizations \cite{cs2}, we will also revisit some of the early ideas and highlight their weaknesses in providing a 
physically viable description. (For a comparative review between old and new quantization, see Ref. \cite{gd}).  Certain properties of the phase space variables will be discussed which prove useful to understand the results from 
quantum theory, in particular whether they can be physically viable. In the last part we will discuss the exactly solvable model (sLQC). These investigations reveal robustness of various results which have been established numerically. In particular, the occurrence of bounce for  states in a dense subspace of the physical Hilbert space, existence of supremum on energy density which turns out to be equal to the critical energy density, various insights on comparison of 
LQC with Wheeler-DeWitt quantization  and the fundamental discreteness of LQC. Further, 
sLQC enables us to show that semi-classicality across the bounce  is preserved \cite{cs1}.

\section{The Classical Phase Space and Kinematics}
We will be interested in the flat $k=0$ model with
spatial manifold  $\Sigma = \mathbb{R}^3$. Since the manifold is non-compact we have to 
fix a fiducial cell ${\cal V}$ to construct the phase space. A simple choice is to consider
a cubical fiducial cell with volume $\ov$ with respect to the fiducial metric $\q_{ab}$:
$V_o = \int_{\cal V}\, \sqrt{\q}\, \d^3\! x$. The classical phase space in LQG is  
in terms of the Ashtekar variables, the SU(2) connection $A^i_a$ and the triad $E^a_i$. Given the 
symmetries of the Robertson-Walker spacetime, these simplify to \cite{abl}
\be\label{AE_defs}
A^i_a \, = \, c  \, \ov^{-1/3} \ow^i_a, ~~~~
E^a_i \, = \,  p \, \sqrt{\q} \, \,  \ov^{-2/3} \, \e^a_i ~
\ee
where $\e^a_i$ and $\ow^i_a$ are the fiducial triad and co-triad compatible with $\q_{ab}$. The
triad $p$ and connection $c$ satisfy
\be \{ c,\, p\} = \frac{8\pi\gamma G}{3} \, .\ee
Here $\gamma$ is the Barbero-Immirzi parameter whose value is determined from the black hole thermodynamics in LQG. The connection and the triad 
are related to the scale factor and its derivative as 
\be
|p| = \ov^{2/3}\, a^2 \,, ~~~ c = \gamma \ov^{1/3} \, \dot a
\ee
(the latter holding on the space of solutions of classical GR only). 

The classical gravitational constraint written in terms of triads and field strength of the 
connection which is given by (with lapse $N = 1$)
\be \label{eq:cgrav}
C_{\mathrm{grav}} = - \gamma^{-2} \int_{\cal
V} \d^3 x\,  \epsilon_{ijk} \,\f{E^{ai}E^{bj}}{\sqrt{|\det E|}}\,
F_{ab}^i ~ \ee
simplifies to $C_{\mathrm{grav}} = - 6 (c^2/\gamma^2) |p|^{1/2}$. Choosing a matter field (such as massless  scalar $\phi$ with momentum $p_\phi$) the  total constraint can be written as
\be
C_{\mathrm{grav}} + C_{\mathrm{matt}} = - 6 \f{c^2}{\gamma^2} |p|^{1/2} + 8 \pi G \, \f{p_\phi^2}{|p|^{3/2}} ~.
\ee
Vanishing of the total constraint and 
solving for the Hamilton's equation for
$c$ we are led to the classical Friedman and Raychaudhuri equations respectively:
\be\label{clfried}
\f{\dot a^2}{a^2} \, = \, \f{8 \pi G}{3} \, \rho ~, ~~~ \f{\ddot a}{a} = - \f{4 \pi G}{3} \, \left(\rho + 3 P \right)
\ee
where $\rho$ is the energy density and $P$ is the pressure.  It is related
to the energy density as $P = w \rho$ where $w$ is the equation of state.
 For the massless scalar, solving above equations we obtain 
$\rho \propto a^{-6}$ which diverges as $a \rightarrow 0$. (We will later see, that LQC leads to an effective Hamiltonian which results in a non-singular modified Friedman dynamics).

The elementary variables used in LQC are the holonomy of the connection $c$ along a straight edge $e$ and the flux integral of the triad involving smearing by a constant test function across a square tangential to the $\e^a_i$. Along an 
edge $\lambda \e^a_k$  with length $\lambda V_o^{1/3}$, the holonomy is given by
\be
h_k^{(\lambda)} = \cos (\lambda \, c/2) \mathbb{I} + 2
\, \sin  (\lambda \, c/2) \tau_k
\ee
where $\tau_k$ are related to the Pauli spin matrices $\sigma_k$ as $\tau_k = - i \sigma_k/2$. The flux integral turns out to be proportional to $p$ up to 
a constant depending on the choice of the cell. 
Elements, $N_\mu:= \exp(i \lambda c/2)$, of holonomies generate an algebra of almost periodic functions of $c$. Using Gelfand construction we can find the representation of this algebra and the kinematical Hilbert space which turns out to be 
${\cal H}_{\mathrm{Kin}} = L^2(\mathbb{R}_{\mathrm{Bohr}},d \mu_{\mathrm{Bohr}})$. Here $ \mathbb{R}_{\mathrm{Bohr}}$ is the 
Bohr compactification of the real line and $d \mu_{\mathrm{Bohr}}$ is the associated Haar measure.

The elements $N_\lambda$ form an orthonormal basis in ${\cal H}_{\mathrm{Kin}}$ and satisfy $\langle N_{\lambda_1}| N_{\lambda_2} \rangle = \delta_{\lambda_1,\lambda_2}$. In the 
${\cal H}_{\mathrm{Kin}}$ the eigenstates of $\hat p$ operator are labeled by
$|\mu\rangle$:
\be
\label{p_act} \hat p| \mu \rangle = \f{8 \pi \gamma \lp^2}{6}
\mu |\mu \rangle ~. ~~
\ee
The holonomy act on kets $|\mu\rangle$ as a shift operator,
\be \widehat h_k^{({\lambda})} |\mu\rangle =
\f{1}{2} \left(|\mu + \lambda\rangle + |\mu - \lambda\rangle
\right) \mathbb{I}
+ \f{1}{i}  \left(|\mu + \lambda\rangle - |\mu -
\lambda\rangle \right) \tau_k ~. \ee
As in the LQG, the strategy is to write the classical constraint in terms 
of holonomies and the triad (flux integrals) and then quantize.  The constraint 
(\ref{eq:cgrav}) consists of two terms. The term involving inverse triad captures the 
aspects of intrinsic curvature and the
other involving field strength  of the extrinsic curvature. 
The inverse triad term can be rewritten 
as
\be \label{cotriad} \epsilon_{ijk}\, \,\f{E^{aj}E^{bk}}{\sqrt{|\det
E|}}\, = \nonumber \sum_k \f{({\rm sgn}\,p)}{2\pi\gamma
G\lambda\, V_o^{\f{1}{3}}}\, \, \mathring\epsilon^{abc}\,\,
\ow^k_c\,\, \Tr\left(h_k^{(\lambda)}\,
\{h_k^{(\lambda)}{}^{-1}, V\}\, \tau_i\right) ~\ee
using the following identities 
of the classical phase space  \cite{thiemanntrick}:
\be
\f{1}{8 \pi G \gamma}\, \{A^d_l, \epsilon^{ijk} \epsilon_{abc} E^a_i E^b_j E^c_k\} = 3 \, \epsilon^{ijl} \epsilon_{abd} E^a_i E^b_j,
\ee
\be ~~ \f{\{A^i_a,V\}}{V^n} = 
\f{\{A^i_a,V^{(1-n)}\}}{(1 - n)}, ~~   e^i_a = \f{1}{4 \pi G \gamma} \, \{A^i_a,V\}~ ,
\ee
where $V = |p|^{3/2}$ is the physical volume of the cell ${\cal V}$.

The field strength term in (\ref{eq:cgrav}) is regulated as in the 
gauge theories.
We consider a square loop $\Box_{ij}$ with sides of length $\lambda V_o^{1/3}$ in the $i-j$ plane 
of the fiducial cell, with the area of the loop shrunk to zero.
 The field strength becomes
\be \label{F} F_{ab}^k\, = \, -2\,\lim_{Ar_\Box
  \rightarrow 0} \,\, \Tr\,
\left(\f{h^{(\lambda)}_{\Box_{ij}}-1 }{\lambda^2 V_o^{2/3}} \right)
\,\, \tau^k\, \ow^i_a\,\, \ow^j_b, ~~~~ h^{(\lambda)}_{\Box_{ij}}=h_i^{(\lambda)} h_j^{(\lambda)}
 (h_i^{(\lambda)})^{-1} (h_j^{(\lambda)})^{-1}\, . \ee
Thus, the gravitational constraint can be written as
\be \label{hc1}
C_{\mathrm{grav}} = \nonumber \lim_{Ar_\square
\rightarrow 0} \,  \sin (\lambda c) \Bigg[-\, \f{1}{2\pi G\g^3}
\f{{\rm sgn}(p)}{\lambda^3}\,  \sum_k \Tr \tau_k
h_k^{(\lambda)}\, \{(h_k^{(\lambda)})^{-1},\, V\}\Bigg] \sin
(\lambda c) ~.
\ee
Due to the underlying quantum geometry, the limit of above operator does not exist
and the loop can be shrunk only to a minimum area. The viewpoint adopted in 
LQC is that this is the minimum eigenvalue of the area operator in LQG:
$\Delta = 2 \sqrt{3} \pi \gamma \lp^2$.\footnote{Recent insights on the area gap revise this value to be equal to twice of above \cite{abhay-wilson}. As expected this only slightly changes some quantitative details in LQC.} The area of the square loop with respect to the physical metric is $\lambda^2 |p|$ which on equating with $\Delta$
leads to $\lambda = \Delta^{1/2}/|p|^{1/2}$. It is then convenient to 
introduce new phase space variables such that the action of holonomies can be
simplified. These are 
\be
\beta = \f{c}{|p|^{1/2}}, ~~ \mathrm{and} ~~ |\nu| = \f{V}{2 \pi \gamma \lp^2}
\ee
satisfying $\hbar \{\b,\nu\} = 2$. The elements of holonomies then become of the
form $\exp(i \lambda_\beta \beta)$ where $\lambda_\beta = \Delta^{1/2}$ is the 
new affine parameter. The corresponding operators have an action of translation on the states $|\nu\rangle$.

\section{Quantum Dynamics}
The quantum operator corresponding to the gravitational constraint 
can be written as
\be
\hat C_{\mathrm{grav}} \Psi(\nu,\phi) = \sin(\lambda_\b \b)
A(\nu) \sin(\lambda_\b \b) \Psi(\nu,\phi)\ee 
where $A(\nu)$ is obtained from the operator corresponding to the 
inverse triad
\be 
A(\nu) = - \f{6 \pi \lp^2}{\gamma \lambda_\b^3}\,
|\nu| ~ \left||\nu + \lambda_\b| - |\nu - \lambda_\b|\right| ~. 
\ee
The action of quantum constraint leads to a quantum difference equation
with uniform steps in volume:
 \be f_+(\nu) \Psi(\nu + 4 \lambda_\b) +
f_0 \Psi(\nu) + f_- \Psi(\nu - 4 \lambda_\b) = \hat C_{\mathrm{matt}}
\, \Psi(\nu) \ee
with 
\be
f_+(\nu) = \f{3 \pi \lp^2}{2 \gamma \lambda_\b^3} \, |\nu + 2 \lambda_\beta| \left||\nu + \lambda_\beta| - |\nu + 3 \lambda_\beta|\right|, ~~ f_0(\nu) = - (f_+(\nu) + f_-(\nu)), ~~ f_-(\nu) = f_+(\nu - 4 \lambda_\b) ~.
\ee
The eigenvalues of $C_{\mathrm{matt}}$ depend on the power of the scale factor
in the classical expression and are modified only if there are inverse powers.
For the massless scalar, we have $1/V$ in the Hamiltonian whose eigenvalues
are given by
\be \widehat{V^{-1}}\,{\Psi}(\nu,\phi) = \f{27}{64}\,
\f{1}{2 \pi \gamma \lp^2 \lambda_\b^3}\, \big|\, |\nu + \lambda_\b|^{\f{2}{3}} - |\nu -
\lambda_\b|^{\f{2}{3}}\,\big|^3\, {\Psi}(\nu,\phi) ~ =: ~ B(\nu) \Psi(\nu,\phi) ~. 
%
\ee
The total constraint operator: $(\hat C_{\mathrm{grav}} + \hat C_{\mathrm{matt}}) \Psi(\nu,\phi) = 0$,
leads to a difference equation which for the massless scalar 
can be casted in the following form:
\be
\partial_\phi^2 \, \Psi(\nu,\phi) = \Theta(\nu) \Psi(\nu,\phi) ~.
\ee
Here $\Theta(\nu)$ is a difference operator in $\nu$ with step size of $4 \lambda_\b$ obtained from the product of $f(\nu)'s$ and the eigenvalues of the inverse volume.
Since there are no fermions in our model, the physical solutions of the quantum constraint are required to be symmetric
under the change of orientation of the triad: $\Pi \Psi(\nu,\phi) := \Psi(-\nu,\phi) = \Psi(\nu,\phi)$. 

The scalar field $\phi$ plays the role of internal clock and the quantum constraint equation can be interpreted as the Klein-Gordon equation in a static spacetime. This leads to the notion of relational dynamics -- the way geometry changes 
with respect to matter. Thus even without having an explicit notion of `time', as in the Path integral methods, we 
can study `evolution'. The quantum constraint superselects a sector, $\epsilon \in [0,4 \lambda_\b)$ and the evolution preserves the lattice $\nu = \epsilon + 4 n \lambda_\b$.

The physical Hilbert space, ${\cal H}_{\mathrm{phys}}$,
 can be found by applying group averaging methods or
demanding the action of Dirac observables be self adjoint.  It
consists of positive frequency solutions of the 
quantum constraint. These satisfy
\be\label{evolve}
- i \, \partial_\phi \, \Psi(\nu,\phi) = \sqrt{\Theta(\nu)} \Psi(\nu,\phi) ~.
\ee
The Dirac observables  of interest are the momentum
of the scalar field and the volume at a given `time' $\phi$
\be
\hat p_\phi \Psi(\nu, \phi) = - i \hbar \f{\partial \Psi(\nu,\phi)}{\partial \phi}, ~~~ |\hat \nu|_{\phi_o} \Psi(\nu,\phi) = e^{i \sqrt{\Theta} (\phi - \phi_o)} |\nu| \, \Psi(\nu,\phi_o) ~.
\ee 
Finally, the physical inner product is given by
\be
 (\Psi_1, \Psi_2)_{\rm phys} =
\f{\lambda_\b}{\pi}\, \sum_{\nu=4n\lambda}\, \f{1}{|\nu|}\,
{\bar{{\Psi}}}_1(\nu,\phi_o)\, \Psi_2(\nu,\phi_o) ~.\ee

We are now equipped to extract predictions from the theory. Given the 
form of the constraint this can be only done via numerical simulations.
The algorithm is to consider a semi-classical state peaked at a classical  
trajectory in a large universe at late times, let us say at $p_\phi = p_\phi^*$ and $\nu|_{\phi_o} 
= \nu^*$, and evolve the state backwards towards the big bang using (\ref{evolve}). As a comparison, we can consider the Wheeler-DeWitt quantum constraint
which can be casted in a similar form as above, except that it is a differential operator in $\nu$.

\begin{figure}[htbp]
\begin{center}
\vskip-0.7cm
\hspace{0.8cm}
\includegraphics[angle=0,width=0.55\textwidth]{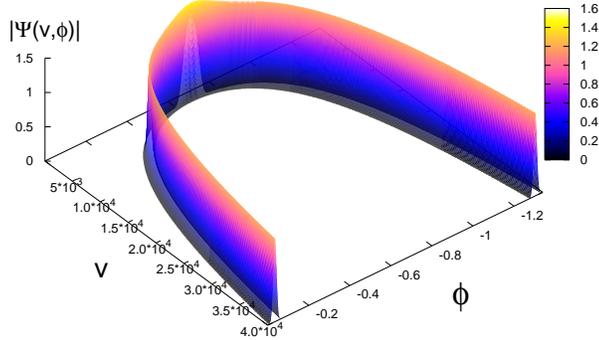}
\caption{The evolution of a semi-classical state is shown in LQC. Unlike the Wheeler-DeWitt quantization the state does not follow the classical trajectory in to the big bang, but bounces to a contracting branch at Planck scale. The lattice chosen in this quantization is  $\epsilon = 2 \lambda_\beta$, $p_\phi^* = 5000$ (in Planck units) and $\Delta p_\phi/p_\phi^* = 0.025$. As can be seen the state remains sharply peaked through out the evolution preserving semi-classicality.}
\end{center}
\end{figure}

Figs 1. and 2, show the result of evolution for such states. The main features are (for details see Refs. \cite{apslett,aps2,apsv}):

\begin{enumerate}
\item States which are semi-classical at late times when evolved backward towards 
the big bang follow the classical trajectory till they reach close to the 
Planck scale. To be precise, the classical theory is an excellent approximation
to LQC till spacetime curvature $R \sim -0.3 \pi/\lp^2$. At higher scales,
departures from GR become significant. At $R = R_{\mathrm{crit}} := -13.12/\lp^2$ ($\rho = \rho_{\mathrm{crit}} := \sqrt{3}/(16 \pi^2 \gamma^3 G^2 \hbar) = 
0.82 \rho_{\mathrm{Planck}}$) the state 
bounces. The quantum bounce is non-singular and to a contracting branch with the same value of $p_\phi$. The big bang singularity is avoided. 
\item In comparison, the states evolved using Wheeler-DeWitt quantum constraint follow the classical trajectory  in to the big bang. Wheeler-DeWitt quantization does not cure the big bang singularity.  
\item States remains sharply peaked through out the evolution in LQC. The relative dispersion of observables remains small before and after the bounce (though they may not be equal). Semi-classicality is preserved across the bounce. 
\item Effects originating from the inverse triad terms in the gravitational and matter constraint turn out to be negligible compared to those from the field strength.
In fact, even if one chooses not to regulate the inverse triad terms which diverge classically at the big bang, one will obtain a very similar evolution  as in LQC for states which are semi-classical at late times i.e. for states which lead to a large classical universe. It is to be emphasized that in the flat model {\it the  inverse triad is not tied to the spacetime curvature}. In fact, no meaningful physics can be associated to the scale at which inverse triad effects become dominant. A reason being that this scale is not independent under the rescaling freedom of the fiducial cell (for details see Appendix B2 of Ref. \cite{aps2}). 
In contrast, the field strength which measures the extrinsic curvature of the spacetime leads to effects occurring at invariant scales.
In the closed $k = 1$ model, since the intrinsic curvature is non-zero, modifications coming from eigenvalues of the inverse triad operator do lead to meaningful physics and an interesting phenomenology, for example a non-inflationary possibility to generate 
scalar invariant fluctuations through thermal mechanisms in the early universe \cite{jm-ps}. 
\begin{figure}[htbp]
\begin{center}
\vskip-0.7cm
\hspace{0.8cm}
\includegraphics[angle=0,width=0.55\textwidth]{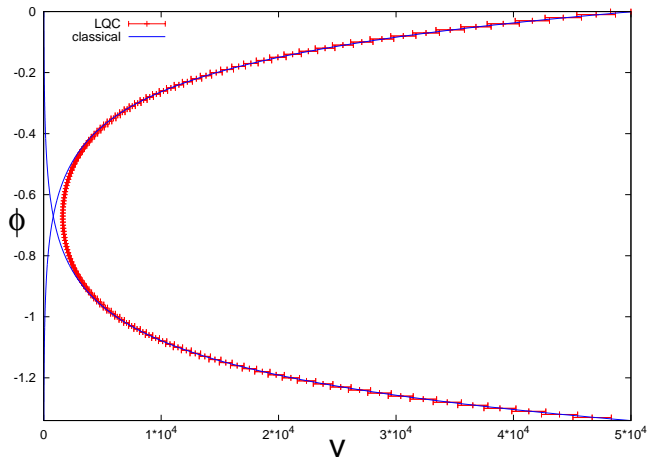}
\caption{Plot of the trajectories from LQC and the classical theory. Classical GR is an 
excellent approximation to LQC till the state reaches Planck scale. Significant departures occur beyond $\rho \sim 0.02 \rho_{\mathrm{Planck}}$. The trajectory from effective Hamiltonian (not shown above) is in excellent agreement with the LQC curve.}
\end{center}
\end{figure}

\item Using geometric methods of quantum mechanics it is possible to write an
effective Hamiltonian which describes the underlying quantum dynamics to 
an excellent approximation. This Hamiltonian is given by \cite{aps2}
\be\label{effham}
\f{3}{\gamma^2} \f{\sin^2(\lambda_\beta \beta)}{\lambda_\beta^2} |p|^{3/2} = 8 \pi G \, H_{\mathrm{matt}} ~.
\ee
The success of the effective Hamiltonian has been extensively tested 
for matter with equation of state $w=1$ (massless scalar) to $w=-1$ (cosmological constant).
Using Hamilton's equations, we can derive the modified Friedman equation\footnote{Interestingly, the modified Friedman equation in LQC has similar structure to the one in  
some braneworld models \cite{sahni}.}
\be\label{fried}
 H^2 = \f{8 \pi G}{3} \, \rho\left(1 - \f{\rho}{\rho_{\mathrm{crit}}} \right)
 \ee
 and the Raychaudhuri equation
  \be\label{rai}
 \f{\ddot a}{a} = - \f{4 \pi G}{3} \, \rho \, \left(1 - 4 \f{\rho}{\rcr} \right) - 4 \pi G \, P \, \left(1 - 2  \f{\rho}{\rcr} \right) ~.
 \ee 
These two equations result in  an unmodified conservation law.
For $\rho \ll \rcr$, the modified Friedman equations reduce to the classical Friedman equations (\ref{clfried}). From the loop quantum modified
Friedman dynamics it is easy to see that when $\rho = \rcr$, the Hubble rate becomes
zero and the universe bounces. In the Planck regime the state, which we evolve backward from a large classical universe, is peaked at the effective trajectory 
obtained from the above equations.
\end{enumerate}

The early quantization in LQC (also known as $\mu_o$ quantization) \cite{abl,mb1}
was lacking in various features as described above \cite{aps1}. The difference was in the 
way field strength tensor is regulated. In the old quantization
 assignment of areas of the loop with respect to the physical
geometry is not considered. Instead, the affine parameter $\lambda$ was assumed to be a constant.
The resulting difference equation was of uniform step in the triad ($\mu$) and 
not in the volume ($\nu$). The quantization predicts a bounce but it occurs
 at a scale which depends on the size of fiducial cell. Thus the scale at which
`quantum gravity' becomes significant can be changed arbitrarily leading to
unphysical effects. Such effects include a generic recollapse of a 
universe at late times when 
dominated by matter which violates strong energy condition \cite{cs2}. (For an exact solution of recollapse see Ref. \cite{polish}). To understand these
issues it is useful to note some features of the phase space variables. \\

\noindent
{\it Underlying freedoms of coordinates and cell:}
For the FRW metric 
\be
\d s^2 = -  \d t^2 + a(t)^2 \, \d {\bf x}^2
\ee
there exists a freedom to rescale the coordinates $x \rightarrow l x$ leaving
the metric invariant. This implies $a \rightarrow l^{-1} a$ and $V_o \rightarrow l^3 V_o$. Under this freedom, the connection and triad are unaffected:
$c \rightarrow c$ and $p \rightarrow p$. 

However, there exists another freedom -- to change only the size of the fiducial cell which amounts to changing the limits of fiducial interval of integration over coordinates: ${\cal V} \rightarrow {\cal V}'$ such that
$\ov' = \alpha^3 \, V_o$. Under this change, 
\be \label{cptrans}
c \rightarrow \alpha c ~~ \mathrm{and} ~~  p \rightarrow \alpha^2 p. 
\ee

\noindent
{\it Variation of phase space variables:}
For a 
general form of matter with a fixed equation of state $w$, the conservation law (obtained from Friedman and Raychaudhuri equation) leads to 
$\rho \propto a^{-3(1 + w)}$, implying
\be\label{conn_scale}
c \, = \, \gamma \dot a \, \propto \, a^{(-3 w + 1)/2} ~.
\ee
Thus for all matter violating strong energy condition $w < -1/3$, the connection increases
as the universe expands. This is different from the behavior of spacetime curvature measured 
for example by the Ricci scalar
which scales as $R \propto a^{-3(1 + w)}$. 

Thus, connection $c$ is neither invariant under the freedom of the choice of the cell nor it faithfully captures the aspects of spacetime curvature. From above properties, it 
is easy to see that the variable $\beta = c/|p|^{1/2}$ which is naturally selected by a consistent 
regularization of the field strength, 
is invariant under various freedoms and also 
scales the same way as the energy density and the curvature. 

Let us now consider the effective Hamiltonian of the old quantization \cite{aps1}:
\be
\f{3}{\gamma^2} \f{\sin^2(\lambda c)}{\lambda^2} |p|^{1/2} = 8 \pi G \, H_{\mathrm{matt}} ~.
\ee
It leads to the modified Friedman equation in the same form as the above 
except that $\rcr$ is not a constant. The value of $\rcr$ can be directly 
obtained from the effective Hamiltonian by computing the energy density 
at which $\sin(\lambda c)$ term saturates. Since 
$c$ is not invariant under the rescalings of the fiducial cell, we find that the saturation
of  $\sin(\lambda c)$ is not independent of $V_o$. This is precisely the reason
for $\rcr$ to depend on $V_o$ and the origin of various unphysical results in the old quantization. As an example, for the massless scalar case $\rcr \propto 1/p_\phi$ which in turn scales with the change in ${\cal V}$.

Using the properties of $\beta$ as noted above, 
a similar argumentation for the effective Hamiltonian (\ref{effham}) 
leads to the conclusion that in the new quantization, bounce occurs at invariant 
curvature scale and there are no departures from general relativity for matter which 
satisfies null energy condition.

{\it Remark:} In literature there are proposals for quantization which rely
on use of variables which are neither $\beta$ nor $c$, motivated from the ideas of lattice refinements \cite{bs1}. Recently, it has been shown that all such proposals are plagued with similar 
problems as in the old quantization of LQC are physically not viable \cite{cs2}. It turns out that for a class of quantizations, only one based on $\beta$ is invariant under freedoms of the choice of fiducial cell and thus lead to quantum bounce at a well defined curvature scale.


\begin{figure}[tbh!]
\begin{center}
\includegraphics[scale=0.6]{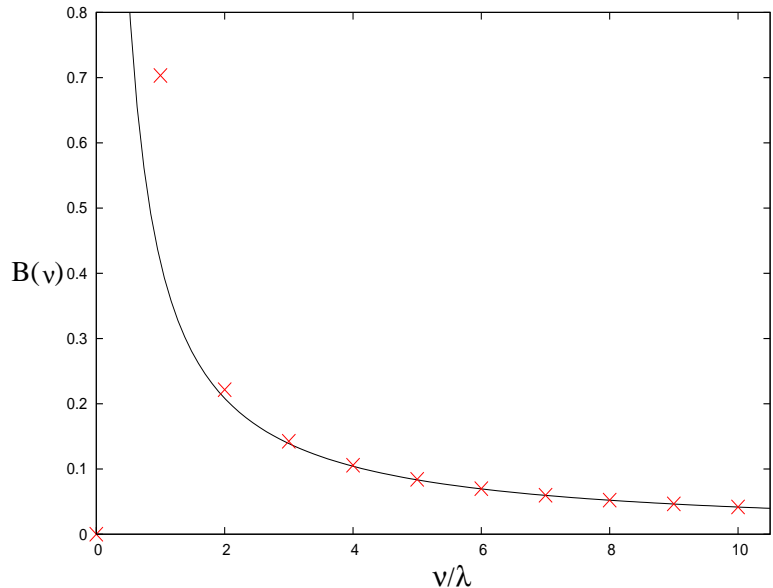}
\caption{Eigenvalues of inverse volume are plotted in LQC (cross) and sLQC (solid). 
For $\nu > \lambda$ the eigenvalues quickly converge to each other. The error is 1.43\%
for $\nu = 4 \lambda$ and 0.02\% for $\nu = 8 \lambda$ respectively.}
\end{center}
\end{figure}

\section{Solvable LQC}
The form of the quantum constraint in LQC makes extracting analytical predictions 
difficult and one has to rely on numerical simulations. However, by choosing 
lapse $N$ to be equal to the physical volume an exactly solvable model can be 
obtained  for the matter content as a massless scalar field \cite{cs1}. This model can also be obtained wth $N=1$ with mild approximations. These are based on the observation that modifications to the constraint originating from the inverse triad play negligible role on the 
singularity resolution and the underlying physics. There are two mild approximations 
involved. Setting $A(\nu) \Psi(\nu) = - 12 \pi \lp^2/(\gamma \lambda_\beta^2) |\nu|$ and 
$B(\nu) = 1/(2 \pi \gamma \lp^2 |\nu|)$ (the Wheeler-DeWitt value). The first approximation is innocuous 
since the expression is exact in LQC for $\nu = 0$ and $|\nu| > \lambda_\beta$.  The second approximation is also
very mild given that departures from the actual inverse volume eigenvalues are extremely
small for $\nu$ as small as $8 \lambda$. The behavior of inverse volume in the 
Wheeler-DeWitt and LQC is shown in Fig 3. As we can see, their departures decrease 
rapidly when we move away from the peak of inverse volume eigenvalues in LQC at $\nu = \lambda$.  

With these approximations one obtains an exactly solvable model in LQC. We emphasize that these approximations are necessary only when lapse is $N=1$. For lapse chosen equal to the volume, exactly solvable model is obtained without any approximation. The Hamiltonian constraint simplifies to
\be
\partial_\phi^2 \,\tilde{\Psi} (\nu, \phi) = 3\pi
G\, \nu\, \f{\sin\lambda_\b \b}{\lambda_\b}\, \nu\,
\f{\sin\lambda_\b\b}{\lambda_\b}\, \tilde{\Psi}(\nu,\phi) 
=: \Theta_{(\nu)}\, \t\Psi(\nu,\phi)\, . \ee
It is also possible to write the constraint in $\b$ representation:
\be \label{hc7}
\partial^2_\phi \, {\chi}(\b,\phi) = 12\pi G\,\, \left(\f{\sin
\lambda_\b\b}{\lambda_\b}\, \partial_\b\right)^2\,\, {\chi}(\b,\phi) \ee
where $\chi(\b,\phi)$ are Fourier transforms of $\chi(\nu,\phi) = (\lambda_\b/\pi \nu) \Psi(\nu,\phi)$. The physical inner product in $\b$ representation is
\be \label{ip4} (\chi_1, \chi_2)_{\rm phys} = \int_0^{\pi/{\lambda_\b}}\,
\d \b\, \bar\chi_1(\b,\phi_o)\, |\hat{\nu}|\, \chi_2(\b, \phi_o) ~.
\ee
We can introduce 
\be\label{x}  x = \f{1}{\sqrt{12\pi G}}\, \ln \left(\tan\left(\f{\lambda_\b\b}{2}\right)\right)
\ee
such that the quantum constraint becomes
\be \label{hc8}\partial^2_\phi\,\, \chi(x,\phi) =
\partial_x^2\,\,\chi(x,\phi) =: -\Theta\,\, \chi(x,\phi)\, .\ee
General solutions can be decomposed in the left moving and right moving components:
 $\chi(x,\phi)= \chi_L(x_+)+ \chi_R(x_-)$. However, the symmetry condition 
$\Psi(\nu,\phi) = \Psi(-\nu,\phi)$ implies $\chi(-x,\phi) = - \chi(x,\phi)$, thus 
\be \label{symmetry} \chi (x,\phi) = \f{1}{\sqrt{2}}\, (F(x_+) -
F(x_-)) \ee
where $F(x)$ are positive frequency solutions of the quantum constraint.

We can also recast the Wheeler-DeWitt theory in the $\b$ representation whose constraint becomes
\be \label{hc1} \partial^2_\phi\,\, \ul\chi (\b,\phi) = - 12\pi G
\,\, (\b\,\partial_\b)^2 \,\,\ul\chi(\b,\phi)\, .
 \ee
Defining 
\be \label{y}  y := \f{1}{\sqrt{12\pi G}}\, \ln \f{\b}{\b_o} 
\ee
the constraint takes a similar form as (\ref{hc8}):
\be \label{hc2} \partial_\phi^2\,\, \ul\chi (y, \phi) =
\partial_y^2\,\, \ul\chi(y,\phi)
=: - \ul\Theta\,\, \ul\chi(y,\phi) \ee
with $\b_o$ a constant. However unlike sLQC, the left and right moving components of solutions of the
Wheeler-DeWitt  constraint are independent of each other. The left and right moving sectors
are further left invariant by the Dirac observables: $\hat p_\phi$ and $\hat V|_{\phi_o}$.
The inner product for the Wheeler-DeWitt quantization can be written as
\be \label{ip2} (\ul{\chi}_1,\, \ul{\chi}_2)_{\rm phy} =
\int_{-\infty}^{\infty} \d y \,\, \bar{\ul\chi_1} (y,\phi_o) \,\,\,
|\!-\!2i\p_y|\,\, \,\ul\chi_2 (y,\phi_o) ~.
\ee

Given the close similarity between the quantum theories of sLQC and Wheeler-DeWitt, it is
necessary to bring out the key difference. It lies in the action of the volume observable.
To illustrate it, let us consider the volume observables in Wheeler-DeWitt theory and without any loss of generality focus on the left moving sector (the expanding branch):
\be \label{wdwvol}(\ul{\chi}_L,\, \hat V|_{\phi} \,
\ul{\chi}_L)_{\mathrm{phy}} =  2 \pi \gamma \lp^2 \,
(\ul{\chi}_L,\,\,|\hat \nu|
\ul{\chi}_L)_{\mathrm{phy}} = \underline{V} \, e^{\co \phi},
\ee
here $\underline{V}$ is a constant
determined completely once the initial data is specified. Thus for any given state, 
the volume observable diverges as $\phi \rightarrow \infty$ and vanishes when $\phi \rightarrow - \infty$. The backward evolution leads to a big bang singularity for all the states in  
Wheeler-DeWitt theory.

The volume 
observable in sLQC yields
\be
 (\chi,\,\, \hat V|_{\phi} \, \chi)_{\mathrm{phy}}=
2 \pi \gamma \lp^2 \, (\chi,\,\,|\hat \nu| \chi)_{\mathrm{phy}} = V_+ \,
e^{\co \phi} + V_- \, e^{-\co \phi} ~. \ee
Here $V_+$ and $V_-$ are positive definite constants determined by the initial data. Unlike
the Wheeler-DeWitt theory the volume observable becomes infinite both in asymptotic future
and past, attaining a minimum volume $V_{\mathrm{min}} = 2 \sqrt{V_+ V_-}/||\chi||^2$ at 
\be
\phi_{\mathrm{bounce}} = \f{1}{(2 \co)} \log(V_-/V_+). 
\ee
Thus, in sLQC  {\it for any state, the backward 
evolution leads to a quantum bounce}. The exactly solvable model enables us to extend the 
results obtained from numerical simulations in LQC using semi-classical states at late time to a dense subspace of the physical Hilbert space. We summarize the main results below:

\begin{enumerate}
\item {\it Critical energy density as the supremum:} We can construct energy density 
observables and consider their expectation values for general states $\chi(x,\phi)$.
It turns out that the energy density has an absolute upper bound in the physical Hilbert space equal to the $\rcr$ observed in the numerical simulations and the effective dynamics of LQC.
\item {\it Issues of semi-classicality:} For a very large class of states, the relative dispersion in observables is preserved across the bounce. These states include the ones with arbitrary squeezing. For more general states, the asymmetry in relative dispersion across the bounce is significantly bounded by the initial dispersion in the conjugate variables. Though the relative dispersion may be different in the expanding and the contracting branch, the resulting states are still peaked extremely well on the classical trajectories. As an example for a universe which grows to the size of a 1 MegaParsec the difference in the relative dispersion across the bounce is bounded by $10^{-56}$.  If one starts with a state which is sharply peaked in the conjugate variables in a large classical universe, one gets a  state which is sharply peaked on the classical trajectory after the bounce. Semi-classicality is preserved to an excellent degree across the bounce \cite{cs1}.\footnote{Contrary to some claims in the literature, this is true even with the exactly solvable model in Ref. \cite{bs2} (after correcting certain dimensional errors). In that analysis a much stronger requirement in the form of dynamical coherence is imposed. For a 1 MegaParsec universe, the resulting  difference in relative dispersion across the bounce is bounded by $10^{-112}$. For a discussion  we refer the reader to Ref. \cite{reply}.}

\item {\it Comparison between sLQC and Wheeler-DeWitt:} This question is tied to the 
possibility of turning off the quantum geometry effects by taking the limit $\Delta \rightarrow 0$. It can be shown that for a given fixed value of $\Delta > 0$ and an 
$\varepsilon > 0$, there exists a finite time interval such that sLQC and Wheeler-DeWitt
approximate each other within $\varepsilon$. However, for a global time evolution the 
predictions of the theory will be drastically different.
\item {\it Fundamental discreteness of sLQC:} As in the case of Wheeler-DeWitt and sLQC, we can compare two sLQC theories with different $\Delta$ parameters. We then find that 
sLQC does not admit a limit when $\Delta \rightarrow 0$. The use of area gap to regulate 
field strength is a necessity in LQC which leads to its fundamental discreteness.

\end{enumerate}


\section*{Summary}
Loop quantum cosmology via the incorporation of non-perturbative quantum gravity effects 
has given useful insights on the quantum nature of the big bang. Its success lies in overcoming the limitations of the Wheeler-DeWitt quantum cosmology. From the studies of 
simplest models the emerging picture resolves the big bang singularity.
The quantum geometric effects lead to significant departures from classical GR at Planck scale leading to a quantum bounce. The spacetime does not end at the big bang singularity, as in classical GR, but extends in to a pre-big bang branch joined with the post big-bang branch 
through a quantum gravitational bridge.\footnote{This is in contrast to models in which singularity avoidance is proposed to occur via the effects of quantum foam at the Planck scale \cite{np}.} The evolution in the Planck regime is fully 
deterministic. Interestingly, for states which correspond to a large classical universe 
at late times, it is possible to write an effective Hamiltonian and obtain modified 
Friedman dynamics which leads to an interesting phenomenology with implications for 
inflation and cyclic models \cite{lqc-phen}.

Lessons from failures of old quantization in LQC and limitations of various
other proposals must be incorporated to develop richer models providing a realistic 
description of our Universe \cite{cs2}. It is pertinent to ask \cite{nkd}: Whether the results of 
singularity resolution are artifacts of the symmetries of the cosmological spacetimes or are more general features of the quantum theory? Is there a non-singularity theorem and a non-singular Raychaudhuri equation in general? Current research in the field is aimed to investigate these issues \cite{others}.

\section*{Acknowledgments}
It is pleasure to thank the organizers for  organizing a wonderful meeting and a warm
hospitality. We are grateful to Abhay Ashtekar, Alejandro Corichi, Tomasz Pawlowski and Kevin Vandersloot for extensive discussions and various collaborations.  Research at Perimeter Institute is supported by the Government
of Canada through Industry Canada and by the Province of Ontario through
the Ministry of Research \& Innovation.

\end{document}